# Robust Ptychographic Reconstruction with an Out-of-Focus Electron Probe

Shoucong Ning[1†*], Wenhui Xu [2†], Pengju Sheng[3†], Leyi Loh[4], Yue Lin[5], Fucai Zhang[3], Stephen J. Pennycook[6], Michel Bosman[7*], Qian He[7*]

1. Department of Physics, University of Science and Technology of China, Hefei, Anhui 230026, People's Republic of China.
2. Paul Scherrer Institute, Villigen PSI, Villigen, Switzerland
3. Department of Electrical and Electronic Engineering, Southern University of Science and Technology, Shenzhen 518055, China
4. Department of Physics, National University of Singapore, 117551, Singapore.
5. Hefei National Research Center for Physical Sciences at the Microscale, University of Science and Technology of China, Hefei, Anhui 230026, P.R. China.
6. School of Physical Sciences and CAS Key Laboratory of Vacuum Physics, University of Chinese Academy of Sciences, Beijing, 100049, China
7. Department of Materials Science and Engineering, National University of Singapore, 117575, Singapore

† These authors are equally contributed to this paper

* Email Address: ningustc@ustc.edu.cn

**Abstract**

As a burgeoning technique, out-of-focus electron ptychography offers the potential for rapidly imaging atomic-scale large fields of view (FoV) using a single diffraction dataset. However, achieving robust out-of-focus ptychographic reconstruction poses a significant challenge due to the inherent scan instabilities of electron microscopes, compounded by the presence of unknown aberrations in the probe-forming lens. In this study, we substantially enhance the robustness of out-of-focus ptychographic reconstruction by extending our previous calibration method (*the Fourier method*), which was originally developed for the in-focus scenario. This *extended Fourier method* surpasses existing calibration techniques by providing more reliable and accurate initialization of scan positions and electron probes. Additionally, we comprehensively explore and recommend optimized experimental parameters for robust out-of-focus ptychography, including aperture size and defocus, through extensive simulations. Lastly, we conduct a comprehensive comparison between ptychographic reconstructions obtained with focused and defocused electron probes, particularly in the context of low-dose and precise phase imaging, utilizing our calibration method as the basis for evaluation.

**Introduction**

Electron ptychography has emerged as a powerful tool for recovering the intrinsic phase distribution of specimens, offering improved signal-to-noise ratio and spatial resolution(Chen et al., 2020, 2021; Jiang et al., 2018; Hao Yang et al., 2016; Wang et al., 2017; Zhou et al., 2020; Song et al., 2019). This technique significantly enhances the capabilities of scanning transmission electron microscopes (STEM) equipped with high-speed single-electron detection cameras capable of recording the full diffraction pattern at each scan position. On one hand, the structural information of materials can be imaged with a much lower electron dose and a wavelength limited spatial resolution compared to conventional STEM imaging techniques. This broadens the application scope of STEM imaging to include high-resolution structural characterization of beam-sensitive materials(Pennycook et al., 2019; Jannis et al., 2021; Pennycook et al., 2020; Hao Yang et al., 2016; Lozano et al., 2018; Song et al., 2019), and even biological samples(Zhou et al., 2020; Pei et al., 2023). On the other hand, accurate correction of residual lens aberrations enables delivery of electromagnetic fields related to material functionalities with high fidelity(H. Yang et al., 2016). For instance, C. Zhen et.al (Chen et al., 2022) reported a high precision measurement of the beam deflection angle using Lorentz electron ptychography.

Compared to in-focus ptychography, the out-of-focus case allows for a larger scanning step size due to the increased probe size on the sample plane(Edo et al., 2013; Huang et al., 2014) (**Fig 1**). As a result, a significantly expanded field of view (FoV) without sacrificing resolution becomes achievable with a fixed number of scanning positions when the electron probe is defocused(Chen et al., 2020; Zhou et al., 2020). Unlike current STEM imaging techniques, ptychography reconstruction does not necessarily adhere to the Nyquist sampling criterion, especially in the out-of-focus scenario. Consequently, cross-scale (nm-μm) imaging with sub-angstrom resolution is potentially attainable through out-of-focus ptychography. The high-throughput structural information obtained will enhance STEM's efficacy as a tool for resolving structure-property correlations in both functional and structural materials, particularly with the integration of deep learning techniques.

Unfortunately, the widespread adoption of this technique is hindered by electron microscope instabilities, such as time-varying lens aberrations and positional errors(Muller & Grazul, 2001; Ning et al., 2022, 2018; Jones et al., 2015). These unknown aberrations and positional errors restrict the accurate initialization of the electron probe and scan positions, essential for current ptychography reconstruction methods. While scan positions and probe functions can be refined during iterative ptychography reconstructions (Hüe et al., 2011; Maiden et al., 2017; Thibault et al., 2009; Thibault & Guizar-Sicairos, 2012) using simulated annealing, cross-correlation and gradient-descent based methods (Dwivedi et al., 2018; Beckers et al., 2013; Ning et al., 2023), reliable initialization of scan position and probe function remains necessary for efficiency and robustness. Several methods have been proposed to determine the uniform rotation between scan positions and the electron camera(Savitzky et al., 2021; Hachtel et al., 2018), and we further proposed the *hybrid method* (Ning et al., 2022) to determine the affine transformation of scan-camera coordinates directly on diffraction datasets without prior knowledge of the imaged sample. However, the application of this method is limited to in-focus 4D-STEM datasets. For out-of-focus 4D-STEM datasets, Hurst *et al.* (Hurst et al., 2010) determines the scan positions by tracking the specimen features in its Ronchigram using the *cross correlation (CC) method*. Nevertheless, as will be demonstrated in our study, this method is sensitive to residual lens aberrations, detector gain distribution, and other factors.

In this study, the *Fourier method*, a subroutine of our *hybrid method*, has been successfully extended to determine the affine transformations of scan-camera coordinates for individual out-of-focus diffraction datasets. Through validation with both simulated and experimental data, we have significantly improved the robustness and accuracy of both iterative and non-iterative ptychography reconstructions using this e*xtended Fourier method*. Importantly, our theoretical analysis highlights that the robustness of ptychographic reconstruction relies not only on the accurate initialization of scan positions and probe functions but also on crucial experimental parameters such as defocus and aperture size. Additionally, we show the impact of defocus values on low-dose ptychography and accurate phase imaging based on our progress in robust electron ptychographic reconstruction of thin specimens.

## Method

During the acquisition of diffraction datasets used for electron ptychography, the electron probe is focused by the probe-forming lens on the sample plane, and it usually scans over the sample in a raster style (**Fig 1** (**a**)). After transmitting the sample, the electron beams fall on the electron cameras in the far field, and diffraction patterns are collected. The two-dimensional scan on the sample and the two-dimensional diffraction patterns form four-dimensional STEM (4D-STEM) datasets. The bright field part of electron diffraction patterns, the Ronchigram (bottom of **Fig 1** (**a**)), is a magnified shadow image of the specimen (Pennycook & Nellist, 2011). A larger defocus of the electron probe usually leads to a smaller magnification, but a wider field of view, of the imaged sample. The scan positions on the sample plane are transformed relative to the camera Cartesian coordinate system (*x, y*) in the far field due to complex geometries and instabilities of the microscope. This transformation includes the uniform (affine transformation) and non-uniform types (scanning distortions, etc.), and here we focus on the affine transformation between the scan and camera coordinate system. In this case, the scan vectors $m_h$ and $m_v$ along the fast (*h*) and slow (*v*) scanning direction, respectively, are able to construct the scan positions with affine transformations, and solving $m_h$/$m_v$ in the (*x, y*) coordinate system is the key to the geometric calibration of 4D-STEM datasets. For the out-of-focus case, the *cross-correlation (CC) method* is widely adopted to determine $m_h$/$m_v$ by tracking the movement of imaged samples in the Ronchigram (**Supporting Materials S2**). However, the accuracy of the *CC method* might be a big concern due to the non-uniform magnification of the sample in the Ronchigram.

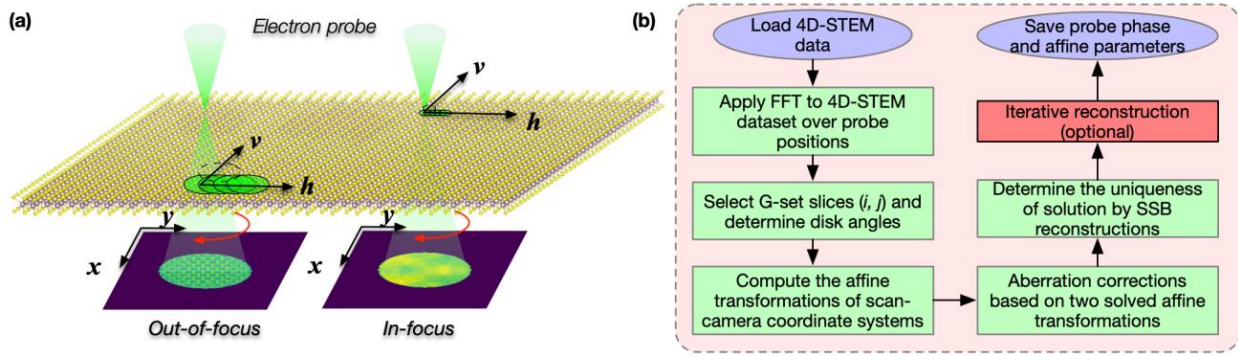

**Fig 1**. (**a**) A comparison of the geometric settings between the in-focus and out-of-focus electron ptychography. In the in-focus case, a focused electron probe scans over the sample and the Ronchigram in the far field shows localized atomic structures. In comparison, a spread electron probe is adopted in the out-of-focus case, and a larger scan step size can be used for ptychography reconstruction. For both cases, the fast scanning direction of the electron probe is defined as *h*, and the slow scanning direction is defined as *v*. Accordingly, the horizontal direction of the electron detector is defined as *x* and the vertical direction is defined as *y*. (**b**) The flowchart for the calibration of out-of-focus 4D-STEM datasets using the *extended Fourier method*.

In our previous work(Ning et al., 2022), the *hybrid method* is proposed to determine the time-varying affine transformation of the scan-camera coordinate system using in-focus 4D-STEM datasets without prior-knowledge about the imaged sample. As the reciprocal-space part of this method, the *Fourier method* can deliver two accurate solutions of the affine transformation with 180° rotation. Then, the unique solution can be confirmed using the real-space *J-matrix method*, which assumes that the nuclei have a positive charge value and higher ADF-STEM intensity when atom columns are resolvable. However, the atom columns are not visible in ADF-STEM images when the probe is out-of-focus, and confirming the unique solution presents a challenge. Taking an experimental out-of-focus 4D-STEM dataset as an example, there are no lattice fringes in the computed ADF-STEM images from this dataset as shown in **Fig 2** (**a**). Fortunately, the *Fourier method* still works since the Bragg peaks still appear in the computed power-spectrum of this 4D-STEM dataset (**Fig 2** (**b**)). The index ($i, j$) of each Bragg peak is only determined by the geometric relationship between scan positions and sample lattices; it has no relationship with the camera placed in the far field. The spatial frequency **Q** of the Bragg peak indexed ($i, j$) is

$$\mathbf{Q} = i\mathbf{k}_h + j\mathbf{k}_v \qquad \textbf{EQ1}$$

where vectors $\mathbf{k}_h$ and $\mathbf{k}_v$ are the reciprocal of scan vectors $\mathbf{m}_h$ and $\mathbf{m}_v$, respectively. In **Fig 2** (**c**), three disks appear in the intensity distribution of the G-set slice corresponding to this Bragg peak, and one of them overlaps with the bright field (BF) disk, and the other two have an opposite shift relative to the BF disk (**Supporting Materials S3**). The shift vector of these two diffracted disks is right **Q** and -**Q**. The shift vector **Q** is only determined by the geometric relationship between the imaged sample and the electron camera, and **Q** can be automatically determined in **Fig 2** (**c**) in our codes. Consequently, $k_h$ and $k_v$ can be uniquely determined when multiple pairs of $(i, j)$ and $\mathbf{Q}$ are provided.

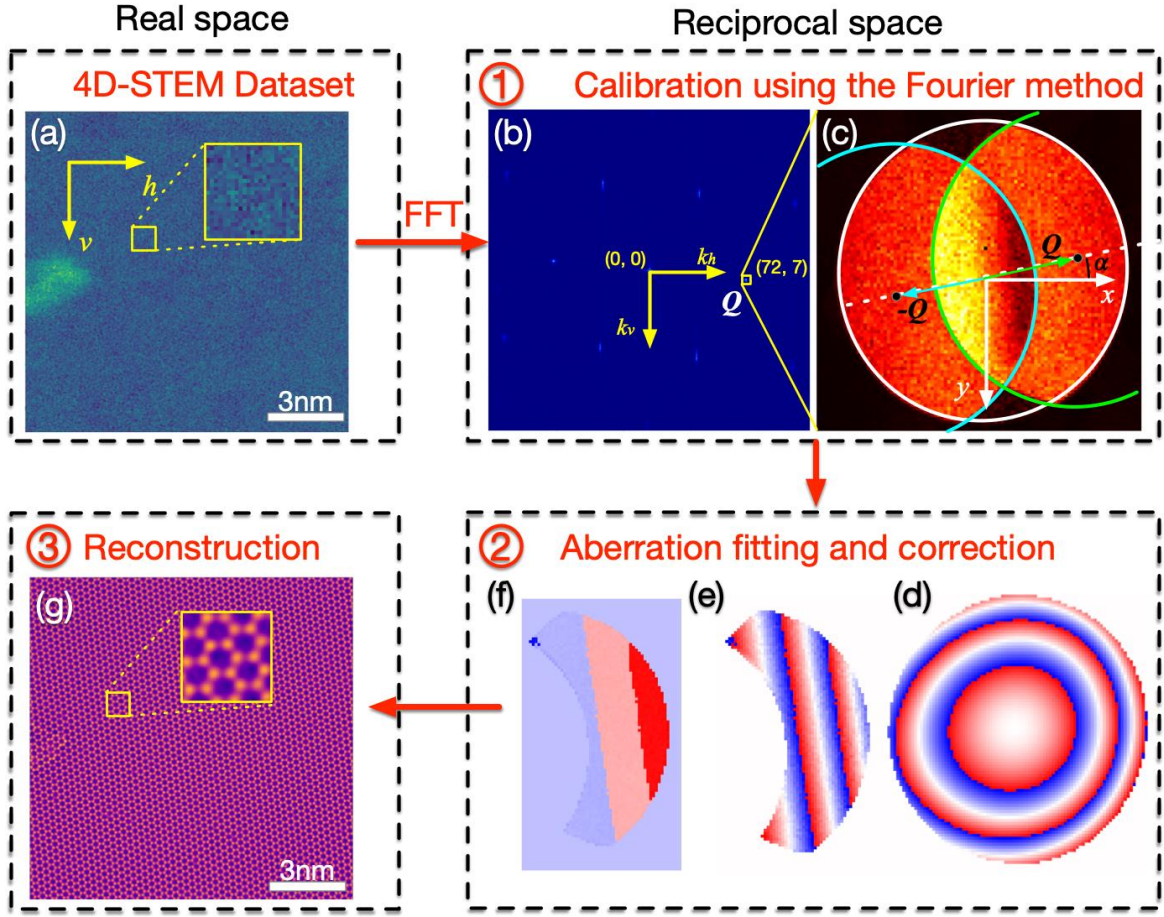

**Fig 2**. The application of the *extended Fourier method* to an experimental out-of-focus 4D-STEM dataset of MoS$_2$ monolayers. (**a**) ADF-STEM image computed from the experimental 4D-STEM dataset. The fast (**h**) and slow (**v**) scanning directions are marked by the yellow arrows in the top-left corner. In the first step, the power spectrum of the 4D-STEM dataset (**b**) is generated after its Fourier transformation. Each pixel of (**b**) corresponds to the total intensity of one G-set slice at the same spatial frequency of the probe, similar to the power spectrum of a conventional atom-resolved STEM image. In comparison, each pixel of (**a**) corresponds to a diffraction pattern at the same scanning position. The unit frequency corresponding to **h** and **v** are $k_h$ and $k_v$, respectively, and the zero spatial frequency of the probe is in the middle of (**b**). The index of each pixel in (**b**) along $k_h$ is $i$ and the index along $k_v$ is $j$. (**c**) is the intensity distribution of the G-set slice with a ($i=72, j=7$) index as marked in (**b**). This G-set slice consists of three disks, the bright field disks and two

diffracted disks, as shown by the white, green and cyan circles in (**b**). The two diffracted disks have the opposite shift vector **Q** and -**Q**, and the intersection angle of **Q** and $k_h$ is α. (**d**) The determined phase distribution inside the aperture using the method given in **Supporting Materials S3** based on the accurate positioning of diffracted disks after geometric calibration. (**e-f**) The phase distribution inside one of the double-overlapped regions in (**c**) before and after aberration correction. (**g**) The reconstructed phase of the MoS$_2$ monolayer after correcting the affine transformation and lens aberrations using the *SSB method*.

Practically, only the rotation angle α of **Q** referring to the horizontal direction of the electron detector *x* is needed. The rotation angles of **h** and **v** relative to *x* and *y* direction, $\theta_x$ and $\theta_y$, as well as γ, the ratio between the modulus of $m_h$ and $m_v$ are solved using the following equation:

$$tan(\alpha) = (i * \cos\theta_y - j * \gamma \sin\theta_x)/(i * \sin\theta_y + j * \gamma \cos\theta_x) \qquad \text{EQ2}$$

At least three Bragg peaks are required to solve $\theta_x$, $\theta_y$ and γ, and these Bragg peaks should have a relatively large intersection angle to ensure robustness. The computed two groups of $\theta_x$, $\theta_y$ and γ values with π rotations right correspond to the **Q** and -**Q** shift of diffracted disks. To further ensure the uniqueness of the solution, the physical meaning of the two diffraction disks should be adopted. When the overlapped regions of the BF disk and the diffracted disk with shift vector **Q** are selected for the phase retrieval of samples, a positive phase distribution is observed around nuclei(Pennycook et al., 2015). When the disk with -**Q** is used for SSB reconstruction, the phase values of nuclei are negative. More importantly, the residual aberrations (**Fig 2 (d)**) of the electron probe, including defocus, should be eliminated in order to reveal the atomic structure of the imaged sample (**Fig 2 (e-g)**). Consequently, the *Fourier method* should be combined with quick ptychography reconstruction and aberration correction to ensure the uniqueness of solutions, and we name this new method as the *extended Fourier method*.

The workflow shown in **Fig 1 (b)** is designed to accurately determine the affine transformation between scan-camera coordinate systems, and ensure the robust iterative ptychography reconstruction of out-of-focus 4D-STEM datasets. In this workflow, the Fourier transformation is first applied to the input 4D-STEM dataset, and three G-set slices with relatively high SNR and large intersection angles are selected to determine the geometric parameters $\theta_x$, $\theta_y$, and $\gamma$. One of the solutions for $\theta_x$, $\theta_y$, and $\gamma$ is used to determine the accurate position of diffracted disks in each G-set slice. Then, the residual aberrations of the probe-forming lens can be robustly determined from one of the double-overlapped regions of BF and diffracted disks with the method given in **Supporting Materials S3**. Based on the accurate position of diffracted disks and the solved aberration coefficient, the *SSB method* based ptychography reconstruction can be successfully applied to unveil the intrinsic atomic structure of the sample. To confirm the correctness of the chosen $\theta_x$, $\theta_y$, and $\gamma$ values, prior knowledge about the imaged sample is required. If the nuclei in the reconstructed phase image have a negative phase value, the other group of $\theta_x$, $\theta_y$, and $\gamma$ is the unique solution and the aberration coefficient of the probe-forming lens should be recomputed. The non-iterative ptychography reconstruction using the *SSB method* is not required to be executed again for the new $\theta_x$, $\theta_y$, and $\gamma$ values, the phase distribution relying on the previous $\theta_x$, $\theta_y$, and $\gamma$ values can be simply reverted and transformed to the coordinate system of the electron camera. At this point, the geometric calibration of the out-of-focus 4D-STEM dataset is completed and the phase distribution of the sample is retrieved (**Fig 2 (g)**). To further eliminate the positional errors of each scan position, and address the incoherence of the electron probe, iterative ptychography reconstruction is suggested. The probe function used for iterative reconstruction, in addition to scan positions, can be initialized from the determined aberration coefficients and geometric parameters, respectively. The reliable initialization of probe function and scan positions leads to the robust and accurate ptychography reconstructions as will be shown in the following experimental results.

## Results

In **Fig 2**, successful ptychography reconstruction has been conducted using the *SSB method* on the experimental 4D-STEM dataset following the workflow given in **Fig 1(b)**. To compare the accuracy of the conventional *CC* and our *extended Fourier* method, the iterative ptychographic reconstruction is conducted and the scanning positions are initialized with the determined geometrical parameters. The determined $m_h$ and $m_v$ values are listed in **Table 1**, and there is 4.5° difference between the intersection angles of solved scan vectors using these two methods. The ePIE method is adopted in the iterative ptychography reconstruction, and two mixed states are used to consider the incoherence of the microscope. The convergence curves are plotted in **Fig 3**

(**a**) for the 2000 ePIE iterations, and the position correction is enabled after 50 iterations to eliminate the non-uniform and residual uniform distortions. The error is the difference between the experimental diffraction patterns and estimated diffraction patterns using computed probe and object functions. As shown, the curve corresponding to the *CC method* has obvious larger errors, and this error slightly increases after nearly 1000 iterations. Therefore, the initialization based on the *extended Fourier method* gives a more stable and faster convergence compared to the *CC method*.

**Table 1**. Calibrated scanning vectors using the *CC* and the *extended Fourier method*.

| Shift vector | $m_h(x, y)$ | | $m_v(x, y)$ | |
|---|---|---|---|---|
| CC method | -0.7807Å | 0.1632Å | 0.0494Å | 0.7961Å |
| Fourier method | -0.7853Å | 0.1395Å | 0.0846Å | 0.7717Å |

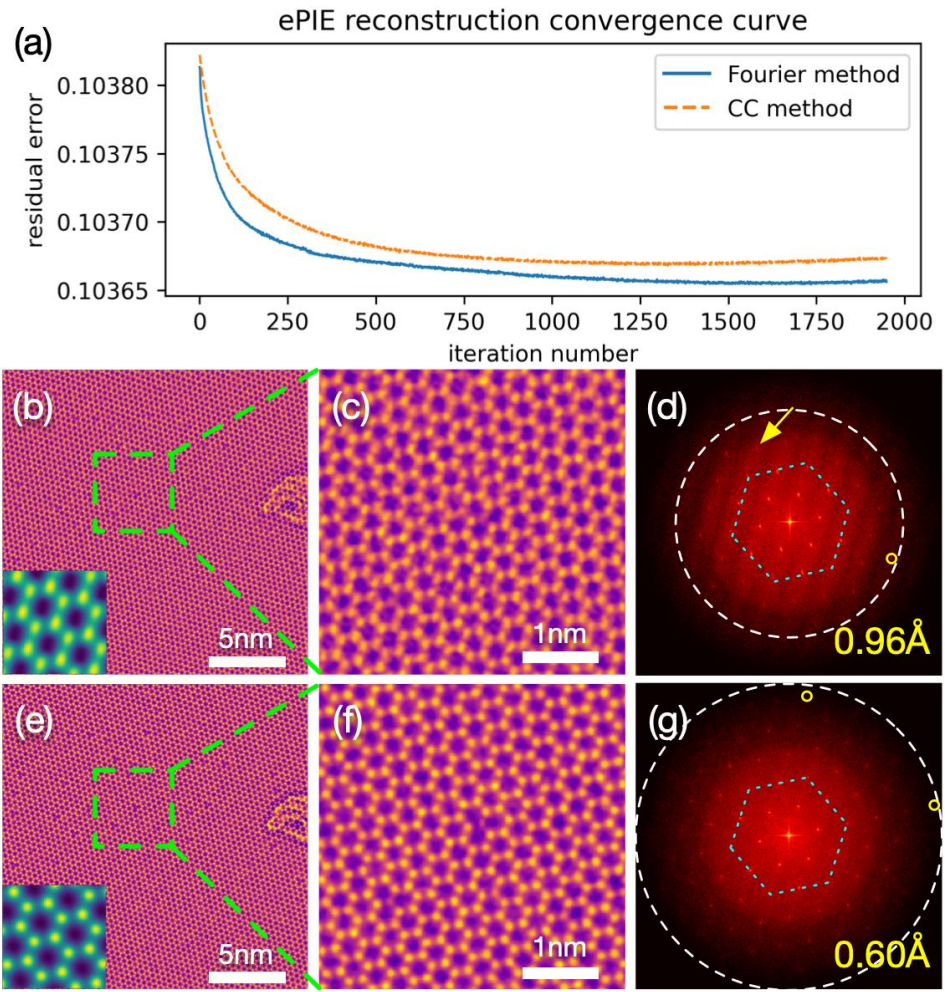

**Fig 3**. Full comparison between the *CC method* and our *extended Fourier method* using the iterative ptychography reconstruction. (**a**). The convergence profiles of the iterative ptychography reconstructions in the 2000 iterations. The error value at each iteration is calculated by normalizing the difference between the estimated and experimental diffraction patterns with the total electron number. (**b-d**) are the reconstructed object phases and its power spectrum, respectively, based on the initialized scanning positions with the *CC method*, and (**e-g**) are results computed based on our *extended Fourier method*. The averaged unit cells generated on (**b**) and (**e**) are inserted into the left-bottom corner of them, and small regions of (**b**) and (**e**) are plotted in (**c**) and (**f**) to show more structural details. Compared to (**b**) and (**c**), the S and Mo atom columns are obviously better resolved in (**e**) and (**f**). In (**d**) and (**g**), the resolution of the reconstructed object is marked by the dotted white circles, and an obvious degradation of resolution is observed in the power spectrum of the reconstructed object based on the *CC method*.

To show more detailed improvements of our *extended Fourier method* over the *CC method*, the reconstructed objects are plotted in **Fig 3** (**b**-**g**). The reconstructed large FoV based on the *extended Fourier method* (**Fig. 3** (**e**)) shows a better contrast compared to the result based on the *CC method* (**Fig 3** (**b**)), and small regions denoted by the green rectangles in **Fig 3** (**b**) and (**e**) are replotted in **Fig 3** (**c**) and (**f**), respectively, to show detailed contrast differences. The gap between the Mo and S columns seems not to be well resolved in **Fig 3** (**c**), and the contrast of the structure is not uniformly transferred after ptychography reconstruction. Further evidence is provided by the averaged unit cells of $MoS_2$ inserted in the left-bottom corner of **Fig 3** (**b**) and (**e**). Notably, the S and Mo atoms almost show the expected centrally symmetric phase distribution in **Fig 3** (**e**) while the symmetry is obviously disrupted in **Fig 3** (**b**).

The power-spectrums of the reconstructed objects are generated in **Fig 3** (**d**) and (**g**) to measure the difference in the symmetry of the reconstructed lattice and resolution. As marked by the yellow arrow in **Fig 3** (**d**), the Bragg peaks are obviously elongated due to the incomplete correction of the non-uniform distortion. While the Bragg peaks in **Fig. 3** (**g**) show sharp contrast, indicating the successful elimination of the non-uniform distortion. Moreover, a dramatic extension of the resolution is observed in the reconstructed result using our *extended Fourier method*, and the measured resolution is around 0.60 Å, while this value is degraded to 0.96 Å due to unreliable initialization of the *CC method*. When connecting the $\{1120\}$ Bragg peaks using regular hexagons, a slight deviation from the Bragg peaks is observed on the vertices of the hexagon in **Fig 3** (**d**), while the 6-fold symmetry is retained in **Fig 3** (**g**). The asymmetry of the reconstructed $MoS_2$ phase based on the *CC method* indicates that ptychography reconstruction requires a reliable initialization of the scanning positions even with the correction of scan positions in the iterative reconstruction process. These experimental results in **Fig 3** are consistent with the simulations given in **Supporting Materials S4**, and current iterative position correction algorithms are not able to correct large uniform deformation of scan positions. Above all, our *extended Fourier method* has a much better performance compared to the existing *CC method* owing to much more accurate initialization.

**Discussion**

*The limits for the probe defocus*

We have shown the effectiveness of our *extended Fourier method*, in addition to the advantage of the out-of-focus 4D-STEM imaging technique. However, the instrumental settings during the experiment are very important for the reproducibility of the out-of-focus ptychography. For example, if the probe defocus is too large, there are no atomic-scale features in the Ronchigram, and then it becomes impossible to retrieve the atomic structure of the object. Consequently, there should be a limitation on defocus value to ensure a successful ptychographic reconstruction. Since the angular value of a single pixel in the electron camera is limited by the camera length settings and the total numbers of pixels, the corresponding box size (red rectangle in **Fig 4** (**b**) ) of the probe in real space is limited. When the probe size exceeds the length of the box, there will be serious aliasing problems, which might produce artifacts in the reconstructed object. For ideal cases with no extra aberrations, the probe size on the sample plane is in a linear relationship with the defocus value. Consequently, the defocus value $\Delta f$ of the probe should be lower than the electron wavelength divided by the product of the aperture diameter $D$ and the angular length of the camera pixels $\alpha_p$ (**Fig 4** (**b**)).

$$\Delta f < \lambda/(D\alpha_p) \qquad \textit{EQ3}$$

Taking the Merlin Medipix3 installed in NUS ARM200CF as an example, the angular size of each pixel is around 0.5 mrad at 8 cm camera length. Given that the wavelength of 80 keV electrons is 0.042Å, then the maximum defocus value with a 23 mrad aperture is around 180 nm. When the camera length is decreased to 6 cm, the defocus value is limited to 120 nm. Practically, the defocus should be even smaller due to the existence of other aberrations.

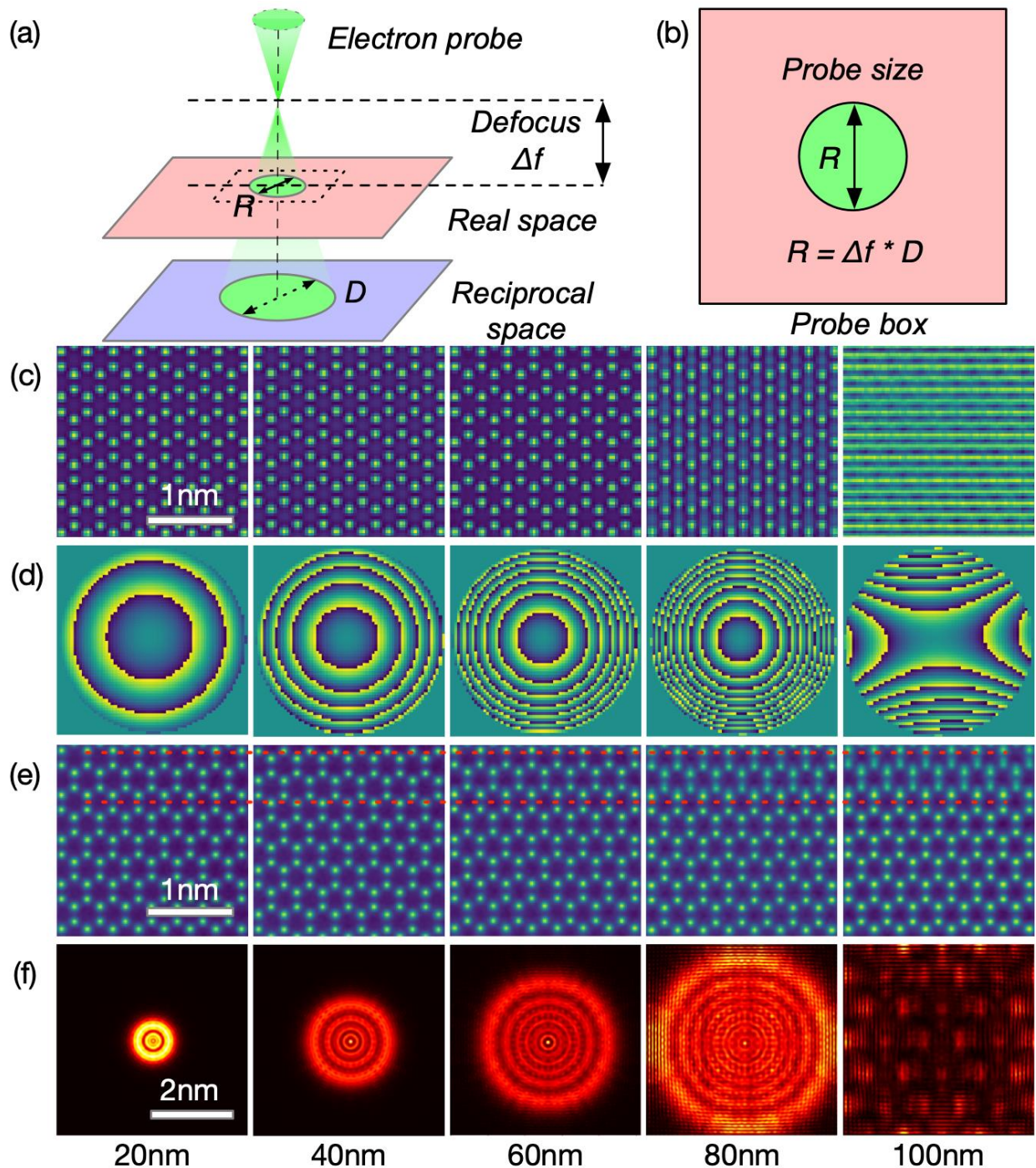

**Fig 4**. The influence of defocus value on the performance of non-iterative and iterative ptychography reconstruction. (**a**) The geometric relationships between the probe size in real space, probe defocus and the reciprocal aperture size. (**b**) To ensure a robust ptychographic reconstruction, the size of the electron probe should be smaller than the probe box, which is determined by the angular value of each detector pixel. In our simulations, the size of the probe box is 44.6 Å, and the corresponding maximum defocus value is around 74 nm for a 30 mrad aperture. (**c**) The reconstructed phase distributions of MoS$_2$ lattices from 4D-STEM datasets with 20 nm, 40 nm, 60 nm, 80 nm, and 100 nm defocus values using the SSB method. When the defocus value is over 74 nm, the reconstructed phase distribution deviates from the intrinsic structure of the MoS$_2$ monolayer. The five images in (**d**) are computed phase distributions inside the aperture of probe-forming lenses from simulated 4D-STEM datasets. An obvious error is observed in the result of the 100 nm defocus case. (**e**) and (**f**) are the computed phase distribution of MoS$_2$ lattices and probe functions using iterative ptychography reconstructions. The atomic structure of MoS$_2$ is retained in all of the five images in (**e**) but unusual contrast appears around 80 nm and 100 nm defocus, as indicated by the red striped line. In addition, an obvious error is encountered in the optimized probe function of the 100 nm defocus case since the obvious symmetry of the MoS$_2$ lattice is observed in (**e**).

To verify the above theory and study the influences of the probe size in more detail, direct simulation is conducted and multiple 4D-STEM datasets with different defocus values are generated. In these simulations, the angular value of each pixel in the virtual detector is around 0.94 mrad, and the aperture size is set to 30 mrad, then the corresponding box size of the probe is 44.6 Å at 80 kV. The computed threshold for the defocus is around 74 nm, and we take 20 nm as an interval to generate the 4D-STEM focal series with infinite doses, and the defocus ranges from 20 to 100 nm. As shown in **Fig 4**. (**c**), the atomic features become distorted in the reconstructed phase distribution using the *SSB method* when the defocus value reaches 80 nm, and atom columns are invisible when the defocus is set to 100 nm. The failure of these reconstructions is due to the inaccurate estimation of aberrations of the probe-forming lens (**Fig 4**. (**c**)). When the defocus is over 74 nm, the angular interval of the probe function in reciprocal space is not enough for the sampling of the phase variation inside the aperture. In the 80 nm case of **Fig 4**. (**d**), two adjacent pixels even have over π phase differences, and the phase unwarping fails when adopting the method in **Supporting Materials S3**. When iterative ptychography reconstruction is applied to these 4D-STEM datasets, the lattice of MoS$_2$ is well resolved with an extended spatial resolution for all of the cases as plotted in **Fig 4**. (**e**). However, fake contrast appears when the defocus exceeds the limit, with the MoS$_2$ atomic structure of the reconstructed area becoming elongated at the top of the images. Although MoS$_2$ lattices are resolved in the 80 nm and 100 nm cases, the reconstructed probe fails, referring to the retrieved probe intensity distributions in **Fig 4** (**f**). Obvious 6-fold symmetry of the MoS$_2$ lattice is even observed in the last image of **Fig 4** (**f**), indicating the failure of iterative ptychographic reconstruction.

*Optimized defocus values for low-dose ptychographic imaging*

For low-dose imaging of beam-sensitive materials and biological samples, the combination of in-focus probes with small scan steps and out-of-focus probes with large scan steps are two typical options(Jannis et al., 2021; Zhou et al., 2020). For the in-focus case, the simultaneously captured ADF signals can be combined with the retrieved phase images from 4D-STEM datasets to give a better understanding of the structural and elemental information of the imaged sample(Wen et al., 2019). An electron camera with a high frame rate is usually necessary in this case since an enormous number of scan positions is required to ensure the overlapping of focused probes. Compared to the focused case, a larger scan step can be adopted for out-of-focus electron ptychography, and the demand for the frame rate is reduced. Theoretically, a defocused electron probe is able to expand the FoV and suppress scan instabilities of ptychographic reconstruction without sacrificing spatial resolution. Currently, which option is most suitable for low-dose structure imaging has not been yet fully discussed.

Taking the low-dose imaging of MoS$_2$ monolayers as an example, extensive 4D-STEM simulations were conducted to evaluate the influence of defocus on the retrieved structural information of MoS$_2$. Twelve 4D-STEM datasets are generated and reconstructed with 0, 20, 40 and 80 nm defocus values and infinite, 1000 e$^{-1}$/Å$^2$, 250 e$^{-1}$/Å$^2$ doses as shown in **Fig 5** (**a**), the aperture size of the probe-forming lens in these datasets is set to 20 mrad (corresponding to a near 110 nm maximum defocus value). To ensure the effectiveness of comparisons between these simulations, the overlap ratios are kept almost the same for all cases by carefully setting the scan step sizes. The focused electron probe has a near 2.0 Å radius after addressing the incoherence; this value increases to 8.0 Å when the defocus is set to 20 nm, and it will be further enlarged to 16.0 Å and 32.0 Å when the defocus is set to 40 nm and 80 nm. Consequently, 0.3484 Å, 1.3946 Å, 2.7892 Å and 5.5784 Å step sizes are used for in-focus, 20 nm, 40 nm and 80 nm cases, respectively. For all of the cases in **Fig 5** (**a**), the reconstruction quality obviously degrades with reduction of the electron dose, but the S and Mo atoms remain visible. An obvious variation of phase background is observed in the in-focus case with a 250 $e^{-1}/Å^2$ dose due to poor transfer of the low-frequency information. To quantify the differences between the reconstructed phase distribution at low electron doses, the positions and amplitudes of all atom columns are used since they are the most important features in structural imaging of beam-sensitive materials. These values are computed for all of frames in **Fig 5** (**a**) using a standard Gaussian model and the deviations of low-dose cases from the values in the infinite dose cases are computed and plotted in **Fig 5** (**b**) and (**c**). In **Fig 5** (**b**), the positional error of atom columns, which is the difference between infinite case and low-dose cases, reduces as the defocus increases to 20 nm, while the error further rises as the defocus changes from 20 to 80 nm, and becomes larger than the in-focus case when the dose is 1000 $e^{-1}/Å^2$. Similarly, the variation of phase first falls and then increases as the electron probe spreads, but the defocus with the minimum phase variation is shifted to the right compared to the positional error. Consequently, the defocus values should be carefully selected for structural imaging of beam-sensitive crystalline samples, and the appearance of residual aberrations in the probe-forming lenses creates structured illuminations for high-quality low-dose imaging.

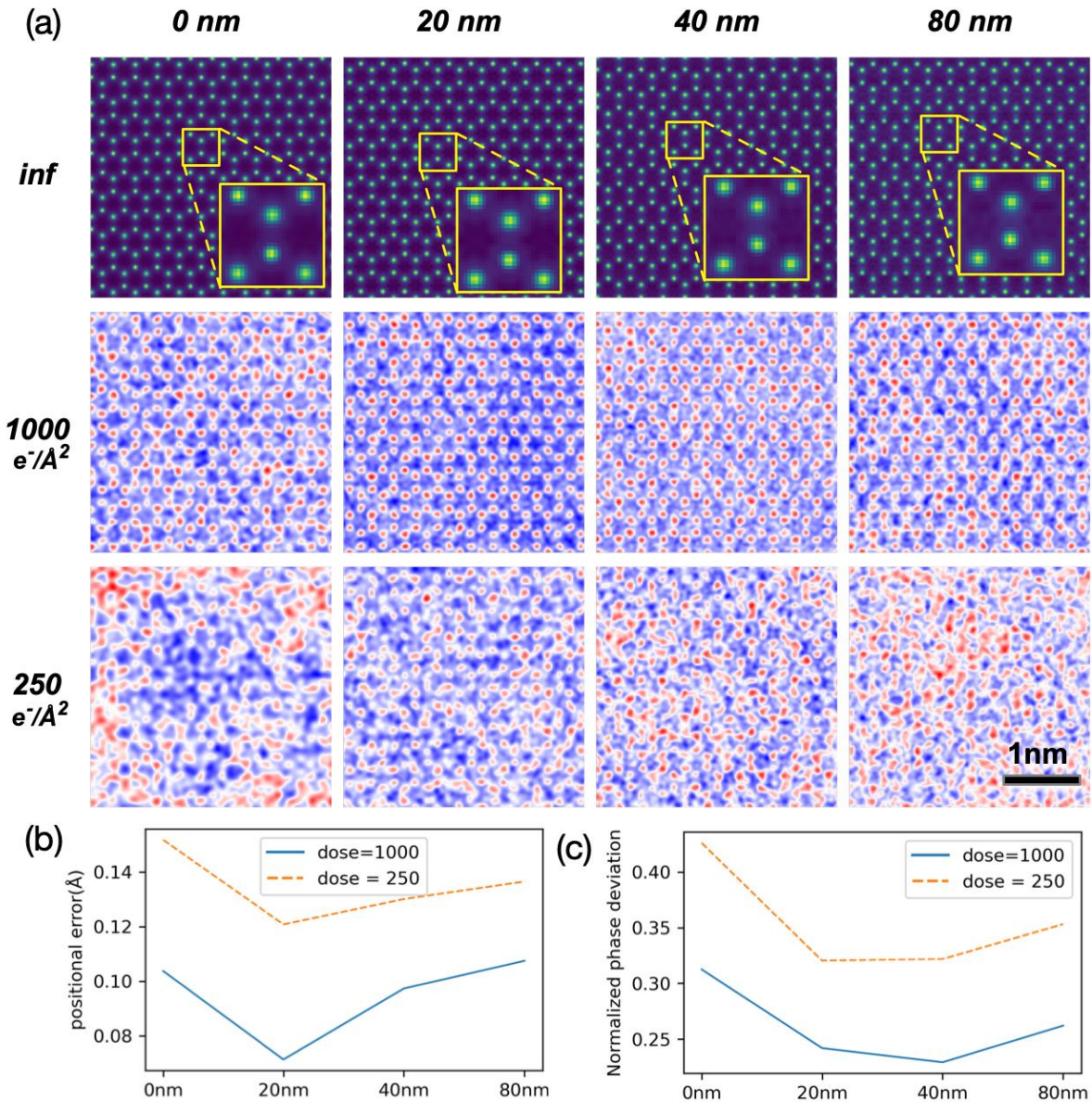

**Fig 5**. The impact of probe defocus on the performances of low-dose electron ptychography. In (**a**), the reconstructed phase distributions of a MoS$_2$ perfect lattice are plotted with the defocus value of the corresponding 4D-STEM datasets set to 0 nm, 20 nm, 40 nm, and 80 nm from left to right, and the electron dose set to infinite, 1000 e$^{-1}$/Å$^2$ and 250 e$^{-1}$/Å$^2$. (**b**) is the averaged positional differences of atom columns at 1000 e$^{-1}$/Å$^2$ and 250 e$^{-1}$/Å$^2$ doses from the infinite dose case at 0 nm, 20 nm, 40 nm defocus values. (**c**) is the normalized phase standard deviations of atom columns at 1000 e$^{-1}$/Å$^2$ and 250 e$^{-1}$/Å$^2$ doses at 0 nm, 20 nm, 40 nm defocus values. Taking the atom column positions at infinite dose as reference, the atom columns at limited electron dose are identified and fitted using the Gaussian model in the generation of (**b**) and (**c**).

*The influence of defocus on contrast transfer*

Although the focused electron probe has slightly poorer performance in low-dose imaging of atomic scale structures, it might have advantages over the out-of-focus case in precise imaging of localized electric fields of materials. In **Fig 5** (**a**), the reconstructed phase distributions with the infinite dose are provided and arbitrary small regions marked by the yellow boxes are amplified in the right bottom corners. As shown, a central symmetric phase distribution of atom columns is observed when the probe is in-focus. As defocus increases, the symmetry is interrupted and apparent 'charge transfer' is observed between the Mo and S atom, but the isolated atom model is adopted in our simulation of 4D-STEM datasets. When the dose is limited, the alterations in the defocus value also result in the emergence of reconstruction artefacts due to the broadened intensity distribution of atom columns. Reconstruction results using simulated 4D-STEM datasets with 0 and 40 nm defocused probes are shown in **Fig 6** (**a**) and (**b**), respectively. The size of the aperture is set to 20 mrad, and the electron dose is set to $1\times10^5$ $e^{-1}/Å^2$. Compared to **Fig 6** (**b**), **Fig 6** (**a**) has a more compact and localized intensity distribution of atom columns when the probe is in-focus. Additionally, **Fig 6** (**b**) exhibits a much stronger background value than **Fig 6** (**a**) with dark spots observed in the middle of the honeycomb-shaped structure of $MoS_2$. A more prominent situation is observed in the experimental results shown in **Fig 6** (**c**) and (**d**) where the electron dose is set to $3.6\times10^5$ $e^{-1}/Å^2$. As the defocus value increases, the electron probe spreads, leading to diffraction patterns contributed by larger sample areas. This diminishes the difference in diffraction patterns across different positions at limited electron dose, subsequently reducing contrast in the reconstruction results. Consequently, more localized information of specimens is transferred when the probe size is small at limited electron doses, and the aberration corrector is still necessary for applications demanding high phase precision, for example, imaging the local electronic structure of 2D materials.

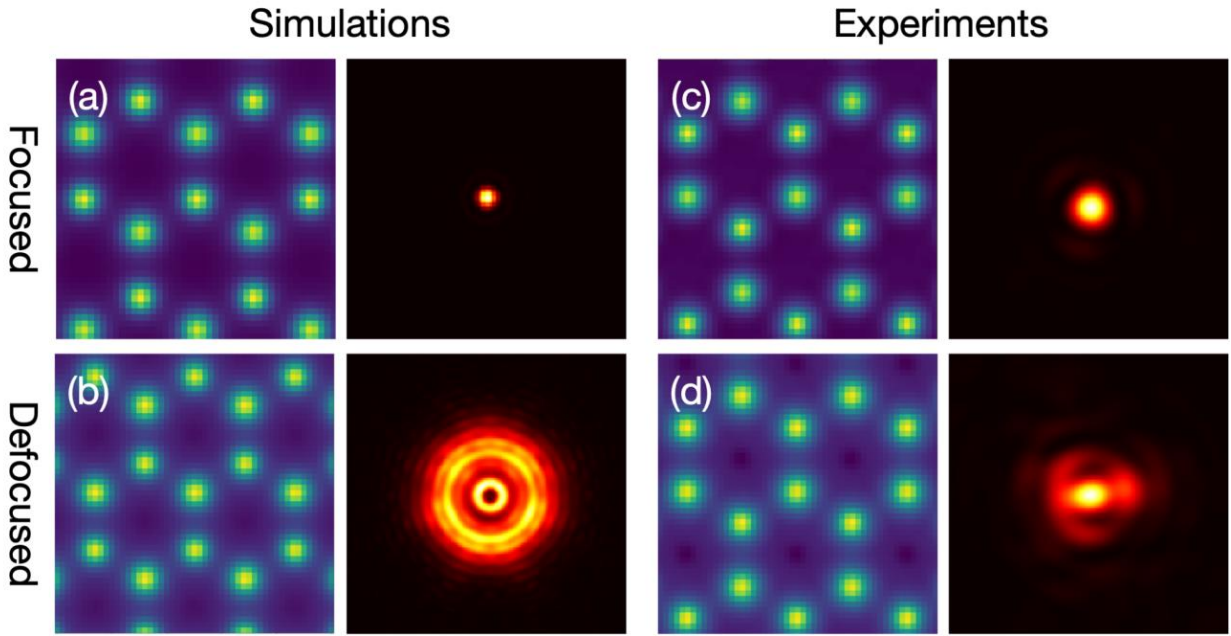

**Fig 6**. The contrast transfer of the ptychography reconstruction with focused and defocused electron probe. (**a-b**) Reconstruction results using simulated 4D-STEM datasets with a focused and 40 nm defocused electron probe. The electron dose is set to $1\times10^5$ e$^{-1}$/Å$^2$. and the aperture size of the probe-forming lens is 20 mrad. (**c-d**) Reconstruction results using experimental 4D-STEM datasets with a focused and 20 nm defocused electron probe, using a 23 mrad aperture and an electron dose of $3.6\times10^5$ e$^{-1}$/Å$^2$. For both simulated and experimental results, the phase distribution around atom columns becomes more dispersed with the increase of defocus values.

**Summary**

In this work, we demonstrate robust out-of-focus ptychography reconstruction of 2D materials by introducing the *extended Fourier method*, and provide optimized criteria for experimental settings of probe defocus. In addition, the impact of defocus on the contrast transfer of localized structural information is studied for both low-dose and high-dose cases. As shown by simulated and experimental results, the out-of-focus ptychography has less advantages in the imaging of localized fine structures compared to the in-focus case. A more important role is played by out-of-focus ptychography in the high-throughput imaging of atomic-scale structures of thin specimens, especially at low-dose. The in-focus ptychography is more suitable for the quantitative study of atomic-scale defects with higher spatial resolution and precision. By combining our *hybrid method* and *extended Fourier method*, robust ptychography can be implemented for both in-focus and out-of-focus cases, enabling a wide application in the structural characterization of thin specimens, such as 2D material monolayers.

## Acknowledgements

M. Bosman gratefully acknowledges support from the Ministry of Education (MOE), Singapore, under AcRF Tier 2 projects MOE2019-T2-1-179 and MOE-T2EP50122-0016. Q. He would like to acknowledge the support of the National Research Foundation (NRF) Singapore under its NRF Fellowship (NRF-NRFF11-2019-0002).